\documentclass{bioinfo}
\copyrightyear{2014}
\pubyear{2014}

\usepackage{url}
\usepackage{graphicx}
\usepackage{multirow}
\usepackage{threeparttable}
\usepackage[multiple]{footmisc}
\usepackage{alltt}

\begin{document}
\firstpage{1}

\title{BEETL-fastq: a searchable compressed archive for DNA reads}
\author[Janin \textit{et~al.}]{Lilian Janin,$^{1}$ Ole Schulz-Trieglaff\,$^{1}$ and Anthony J. Cox$^{1,}$\footnote{to whom correspondence should be addressed}}
\address{$^{1}$Computational Biology Group, Illumina Cambridge Ltd., Chesterford Research Park, Little Chesterford, Essex CB10 1XL, United Kingdom\\}

\address{Computational Biology Group, Illumina Cambridge Ltd., Chesterford Research Park, Little Chesterford, Essex CB10 1XL, United Kingdom}

\history{Received on XXXXX; revised on XXXXX; accepted on XXXXX}

\editor{Associate Editor: XXXXXXX}

\maketitle

\begin{abstract}

\section{Motivation:}
FASTQ is a standard file format for DNA sequencing data which stores both nucleotides and quality scores. A typical sequencing study can easily generate hundreds of gigabytes of FASTQ files, while public archives such as ENA and NCBI and large international collaborations such as the Cancer Genome Atlas can accumulate many terabytes of data in this format. Text compression tools such as gzip are often employed to reduce the storage burden, but have the disadvantage that the data must be decompressed before it can be used.

Here we present BEETL-fastq, a tool that not only compresses FASTQ-formatted DNA reads more compactly than gzip, but also permits rapid search for $k$-mer queries within the archived sequences.
Importantly, the full FASTQ record of each matching read or read pair is returned, allowing the search results to be piped directly to any of the many standard tools that accept FASTQ data as input.

\section{Results:}
We show that 6.6 terabytes of human reads in FASTQ format can be transformed into 1.7 terabytes of indexed files, from where we can search for 1, 10, 100, 1000, a million of 30-mers in respectively 3, 8, 14, 45 and 567 seconds plus 20 ms per output read.
Useful applications of the search capability are highlighted, including the genotyping of structural variant breakpoints and ``\emph{in silico} pull-down" experiments in which only the reads that cover a region of interest are selectively extracted for the purposes of variant calling or visualization.

\section{Availability:}

BEETL-fastq is part of the \texttt{BEETL} library, available as a github repository at \url{git@github.com:BEETL/BEETL.git}.

\section{Contact:} \href{acox@illumina.com}{acox@illumina.com}

\end{abstract}

\section{Introduction}\label{sec:intro}

Much has been written about disruptive changes in DNA sequencing technology over the last decade and the need for compact ways to store the vast datasets that these new technologies have facilitated. The raw reads in a sequencing project are commonly stored in an ASCII-based format called FASTQ (\cite{Cock2010}), the entry for each read comprising a \emph{read ID} string that holds free-text metadata associated with the read, a string for the sequence itself plus a string of \emph{quality scores} that encodes accuracy estimates for each base.

The gzip text compression tool\footnote{\texttt{www.gzip.org}, Jean-Loup Gailly and Mark Adler.} is free and widely available and hence an appealing option for compressing FASTQ files. Many bioinformatics tools can read gzip-compressed data directly via the API provided by the zlib library\footnote{\texttt{www.zlib.net}, Jean-Loup Gailly and Mark Adler.}, which avoids the need to create the uncompressed file as an intermediate, but nevertheless still incurs the computational overhead of decompressing the entire file, which is considerable for the large file sizes we are typically dealing with. For many applications we would like \emph{random access} to the data without the need to decompress the file in its entirety.

This need was recognised by the authors of the Samtools package (\cite{Li2009}), which features two programs, razip\footnote{A tool available from \texttt{razip.sourceforge.net} appears to have similar aims to the eponymous Samtools program, but seems to be an entirely separate development. We deal exclusively with Samtools' \texttt{razip} here.} and bgzip, that take a \emph{block-compressed} approach to random access while retaining some degree of backward compatibility with gzip. The data is divided into contiguous blocks which are compressed individually, allowing decompression to commence from any point in the file, while limiting the overhead to the need to decompress the data that precedes that point within its block. 

Given this functionality, it is simple to index a set of records by some key of interest by sorting them in order of that key, block-compressing them and retaining the mapping from key value to file offset for a subsampling of the records. Several of Samtools' user-level tools work in this way, \emph{e.g.} tabix and indexed BAM files are both indexed by genomic coordinate while faidx uses the name of the sequence as a key.

However, we wish to search for any substring within the sequences, which is not possible with a key-based indexing strategy. Instead, we use an approach based on the Burrows-Wheeler transform, or BWT (\cite{bwt94}). The BWT is a reversible transformation of a string that acts as a \emph{compression booster},  permuting the symbols of the string in a way that tends to enable other text compression techniques to work more effectively when subsequently applied. When the BWT-transformed string is decompressed again, the reversible nature of the transform allows the original string to be recovered from it. 

Although it was originally developed with compression in mind, \cite{Ferragina:2000} showed that, in combination with some relatively small additional data structures, the (compressed) BWT can act as a \emph{compressed index} that facilitates rapid search within the original string. This concept has been highly influential in bioinformatics, being the means by which BWT-based aligners such as BWA (\cite{bwa2009}) and Bowtie (\cite{bowtie2009}) accelerate searches against a reference genome.  The core idea is that exact occurrences of some query within the original string can be found by applying a recursive \emph{backward search} procedure to its BWT. 

Having found some exact matches to our query within the reads in this way, we continue the recursion to extend these hits into the entire sequences of the reads that contain them. Once the extension reaches the boundary of a read, a lookup into an additional table allows the original positions of the reads in the FASTQ file to be deduced. This last piece of information enables the quality score and read ID strings to be extracted from razip-compressed files that have been indexed using their ordering within the original FASTQ file as a key.

Specifically, given a query DNA sequence $q$ our tool can provide, in increasing order of computational overhead:
\begin{itemize}
\item The number of occurrences of $q$ in the reads.
\item The full sequence of each of the reads that contain $q$.
\item The quality score strings associated with the sequences that contain $q$.
\item The read IDs of the reads whose sequences contain $q$.
\item (For paired-read data) The read IDs, sequences and quality scores of the reads that are paired with the sequences that contain $q$. 
\end{itemize}

\section{Methods}\label{sec:methods}

\subsection{Definitions}

Given a string $S$ comprising $n$ symbols drawn from some alphabet $\sigma$, we mark its end by appending a unique symbol $\$$ that is lexicographically smaller than any symbol in $\sigma$. The \emph{Burrows-Wheeler transform} of $S$ is defined such that the $i$-th symbol of $\textrm{BWT}(S)$ is the character of $S$ immediately preceding the suffix of $S$ that is $i$-th smallest in lexicographic order. The concept of the BWT can be readily generalized to encompass a set of strings $S_1, \ldots, S_m$ if we imagine the strings are terminated with distinct end markers that satisfy $\$_1 <\cdots < \$_m$. 

The definition of the BWT implies that any occurrence of a symbol within $\textrm{BWT}(S)$ has a one-to-one relationship with a suffix of $S$ that we call its \emph{associated suffix}. Given some query string $Q$ that occurs at least once in $S$, the characters in $\textrm{BWT}(S)$ whose associated suffixes start with $Q$ form a contiguous subsequence that we call the \emph{$Q$-interval} of $\textrm{BWT}(S)$. If we have located the $Q$-interval in $\textrm{BWT}(S)$, the \emph{backward search} procedure allows the position and size of the $pQ$-interval to be deduced for any symbol $p$ by means of $\mathrm{rank}()$ computations which count the number of occurrences of $p$ within intervals of $\textrm{BWT}(S)$ (see, for example, \cite{bookBWTAdjeroh:2008}).

\subsection{Index construction}

FASTQ data is first split into its three component streams: bases, read IDs and quality scores. The read IDs and quality scores are each dealt with in the same way: compressed with razip and augmented with an index that, for every 1024th read, stores the offset in the file at which the data associated with that read begins. The BWT of the sequences is built using the algorithm described in \cite{BauerCoxRosoneCPM11, BauerCoxRosoneTCS2012}: we use our own BEETL library for this, but we also note the several additional improvements in Heng Li's implementation \footnote{\texttt{github.com/lh3/ropebwt}} and promising recent work by \cite{Liu2014} that demonstrated accelerated BWT construction using GPU technology. 

During BWT construction, we also generate an ``end-pos" file containing an array that maps between the ordering of the read associated with each $\$$ sign and its read number in the original FASTQ file. The BWT itself is stored in a manner similar to that used by the sga assembler (\cite{Simpson2011}), runs of characters being represented by byte-codes, the least significant bits encoding the character and the remainder of the byte denoting the length of the run. To speed the $\mathrm{rank}()$ calculations needed for backward search, we create a simple index of the BWT files by storing, once every 2048 byte-codes, the number of times each character has been encountered so far in the BWT.

The index construction flow is shown in the top half of  Figure~\ref{fig:BEETLFastqFlow}, and Table~\ref{fig:FileTypes} summarizes the files generated by BEETL-fastq to store and index the original FASTQ files.

\subsection{Searching for a query sequence in the index}

\begin {figure}
{\scriptsize    %
$$          %
\includegraphics[bb = 0 0 539 695,width=80mm]{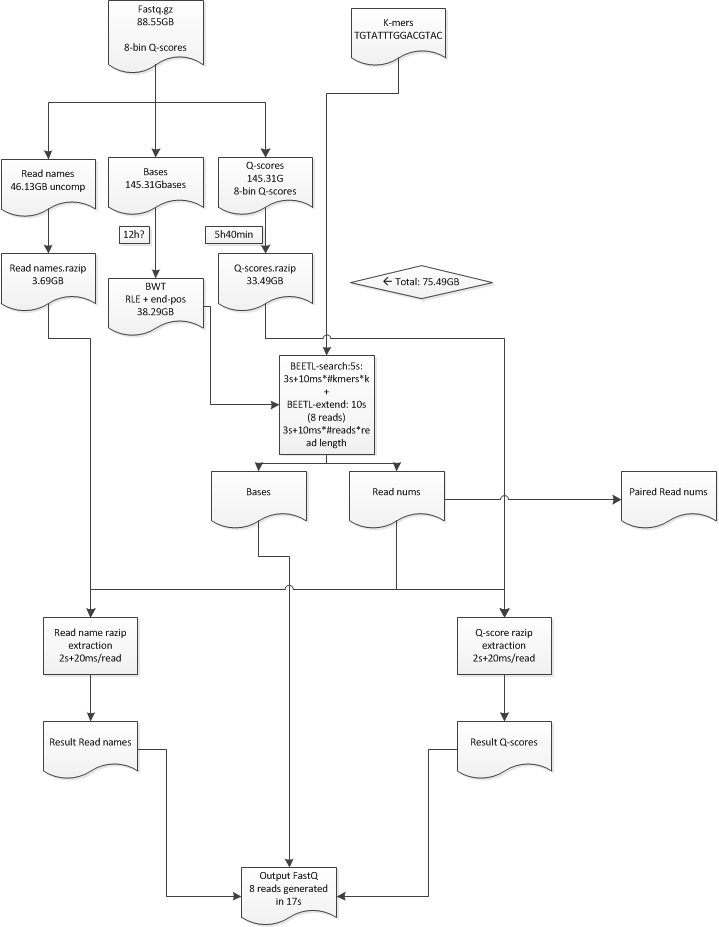}
$$
}
	\caption{BEETL-fastq flow}
	\label{fig:BEETLFastqFlow}
\end{figure}
BEETL-fastq's search mode consists of three main computation stages shown in the lower half of Figure~\ref{fig:BEETLFastqFlow}:
\begin{itemize}
\item BEETL-search performs the k-mer search and retrieves the $Q$-interval matching each k-mer.
\item BEETL-extend propagates each BWT position from these $Q$-intervals to the end of each read, where we are able to identify the read number. At the same time, beetl-extend's propagation of the k-mers reconstructs each base of the reads.
\item From the read numbers, we extract read IDs and Q-scores using the razip files.
\end{itemize}

The read IDs, bases and quality scores are then interleaved to generate the output FASTQ file.

\begin{table}
    \begin{tabular}{|lp{5cm}|}
    \hline
    File name              & Description                                                                        \\
    dataset-B0[0-6]        & BWT files in run-length encoded format                                     \\
    dataset-B0[0-6].idx    & BWT index files for faster random accesses                                         \\
    dataset-end-pos        & Read numbers of the '\$' signs in BWT                                              \\
    dataset.quals.rz       & razip-compressed quality scores, able to start decompressing at any given position \\
    dataset.quals.rz.idx   & index file from compressed quality scores, to map read numbers to character positions in the razip file            \\
    dataset.readIds.rz     & razip-compressed read IDs                                                          \\
    dataset.readIds.rz.idx & index file for compressed read IDs           \\
 \hline
    \end{tabular}
 \caption{Files generated by the indexing process. Together, these files comprise a lossless representation of the data in the original FASTQ files.}
 \label{fig:FileTypes}
\end{table}

\subsection{Cloud-based deployment}

One interesting potential application is as a search service, answering the queries of users on demand and saving them from the need to download very large FASTQ files in full.

A range of use cases were explored, from very fast search of a few $k$-mers to longer batch search of millions of $k$-mers.
Here we report on the two extreme cases: large queries of millions of k-mers under 1 hour, and queries of a single k-mer expected to return results under 1 second, this last use case leading to a useful cloud configuration.
Those timings are achieved on an Amazon EC2 instance querying indexes of the set of 17 human genome datasets described in Section~\ref{subsec:platgenomes}, the indexes being held in Amazon S3 storage.

Both use cases answer the following constraints in different manners:
\begin{itemize}
\item how to efficiently access the datasets stored on S3,
\item what level of parallelisation is needed,
\item how to minimise the costs.
\end{itemize}

Keeping instances running helps achieve good speed, as the data can be kept local, even in RAM if necessary. But this usually comes at a high cost.
On the other hand, spawning Amazon instances on demand takes a few minutes just to start the service.

\subsubsection{Large batch queries}

Large searches of millions of k-mers are actually simple, as the time needed to spawn a large Amazon instance and download the data are small enough compared to the computation time.

Running a cluster of 17 Amazon instances (the number of genomes in our dataset) with 120GB of SSD and 30GB of RAM such as m3.2xlarge satisfies our constraints.
The instances achieve good overall download speed from S3, each being able to keep one full genomes files on disk and its BWT in RAM.
Keeping the genome searches on separate instances prevents RAM contention.

\subsubsection{Small queries with fast response}

Ideally we would like a search for one k-mer within our dataset to yield a response within one second.
However, in practice this timing greatly depends on the number of matching reads as the final razip extraction takes around 20ms per read. We decided to distinguish between the three phases of a search: k-mer search in BWT, extension of BWT positions to bases and read numbers, and extraction of qualities and read IDs.
We focus here on the first stage, the k-mer search, which returns the number of matching reads, an important information which is useful by itself for example in genotyping applications.

The first challenge was to decide where the tools can run: an Amazon instance takes at least one minute to start up, and downloading one of the human genome datasets from S3 to an instance takes in the best case 3 minutes for the sequences alone (around 16GB, assuming 100MB/s).
One solution is to prefetch the BWT files local to an always-on Amazon instance. However this is an expensive solution if the Amazon instance is large enough to process 17 human genome datasets in parallel.

A cheaper alternative is to leave the BWT files in Amazon S3 storage and access them remotely, transparently mounted as files using httpfs\footnote{\texttt{sourceforge.net/projects/httpfs}}.
This is a high latency solution, which ends up working well.
The index files are prefetched and kept in RAM, and the quantity of RAM needed is low enough that we can now pick a much cheaper instance (such as m3.medium).

The organisation of our file system is such that the files described in Table~\ref{fig:FileTypes} appear in one working directory. Large files accessed via httpfs are mounted in sub-directories and symbolic links make them appear correctly in the working directory. Small index files are directly stored in the working directory.
With the default BEETL-search tool, search queries of a single 30-mer were at this stage answered within 3 seconds.

An extra optimisation was necessary to bring this down to 1 second: instead of loading the index files from disk to RAM at the start of each BEETL-search, a shared memory structure keeps the index data permanently in RAM, with the ability to be re-used across successive BEETL-search processes.

In order to make these results easily reproducible and for easy discovery of our tools architecture, a public Amazon AMI image was created, containing all the tools described here: BEETL tools, pre-fetched index files and httpfs remote mounts of the BWT files for the NA12877 human genome sample from S3.
The image also includes a web server able to launch k-mer queries on NA12877.
This image appears as "BEETL-fastq Bioinformatics" in Amazon AMI search.

\section{Results}\label{sec:results}

\subsection{The Platinum Genomes dataset}\label{subsec:platgenomes}

In the following sections we refer to several data sets from Illumina's Platinum Genome (PG) project (Eberle \emph{et al.}, manuscript in preparation), which aims to systematically identify all variants 
in a large three-generation family (the CEPH/Utah pedigree 1463) by combining multiple variant calling approaches and making use of the inheritance structure within the pedigree. Both the raw sequences and variant calls are publicly available.\footnote{\texttt{www.illumina.com/platinumgenomes/}} 
The total dataset comprises 27 billion reads of length 101, requiring 6.6TB of storage in FASTQ format (or 2.5TB if compressed with gzip).

Our experiments with these genomes involved repeated searches of k-mers.
Our strategy was to prepare and search each genome independently to get the benefits of distributed processing.

Each of the 17 datasets is a human genome sequenced at a 50x coverage depth. They are therefore similar in size.
One of them, the NA12877 dataset contains 818,908,462 paired reads of length 101, or 165.4 billion bases.
It comprises 8 fastq.gz files totalling 151.71 GB, equivalent to 390.55 GB of uncompressed FASTQ data.
Compressing and indexing this data with BEETL-fastq (\emph{i.e.} conversion of gzipped FASTQ into BWT plus razip-compressed quality scores and read IDs) took 7.5h on an Amazon EC2 i2.2xl instance and reduced the data to 113GB, a reduction of 26\%.

\begin{table}
    \begin{tabular}{|llllll|}
\hline
    Number of 30-mers             & 1               & 10  & 100  & 1,000 & 1,000,000 \\
    beetl-search (30 cycles)  & 1               & 1.8 & 2.9  & 7.2   & 207       \\
    beetl-extend (100 cycles) & 1.4             & 6.1 & 11.2 & 37.5  & 360       \\
\hline
    \end{tabular}
 \caption{BEETL-fastq search speed from NA12877 (time in seconds)}
 \label{fig:BeetlFastqTimings}
\end{table}

From this, search query results were obtained at a rate presented in Table~\ref{fig:BeetlFastqTimings}, while the subsequent read extraction stage consistently took around a further 20 milliseconds per read.

\subsection{Reads re-ordering}

As an opportunity for optimisation, we note that re-ordering the original reads according to different strategies affects the compression rate of the generated files and the computation speed for both compression and search.

We experimented with 3 strategies:
\begin{itemize}
\item unordered, where the reads are left in the original FASTQ order
\item Lexicographically Ordered (LO) sort, where reads are sorted in lexicographical order of their bases
\item Reversed Lexicographically Ordered (RLO) sort, where reads are sorted in lexicographical order of their reversed (not reverse-complemented, although it would be equivalent) string.
\end{itemize}

\begin{table}
    \begin{tabular}{|p{38mm}lll|}
    \hline
    ~                                                & unsorted & LO sort & RLO sort \\
       NA12877-B0* (GB)                                     &    33  &   29.4 & 18     \\
       NA12877-end-pos (GB)                                 & 7.6     & 0       & 7.6     \\
       NA12877.qual.rz* (GB)                                & 64.5     & 64    & 64.1     \\
       NA12877.readIds.rz* (GB)                             & 7.9     & 14.9    & 14.9     \\
       total (GB)                                    & 113     & 108.3    & 104.6     \\
       reduction vs fastq.gz                      & -26\%     & -29\%   & -32\%     \\
       compression time   (min)                     & 590        & 602      & 647        \\
       search-only time for one 30-mer (ms) & 505       & 460      & 280        \\
       \hline
    \end{tabular}
 \caption{Effect of sorting strategies. Dataset: Platinum Genome NA12877}
 \label{fig:SortingStrategies}
\end{table}

Table~\ref{fig:SortingStrategies} shows the results obtained on Platinum Genome NA12877.
The compression and search times were obtained on a server with 16 2.5 GHz cores, although BEETL-fastq is using at most 6 cores. BWT construction is I/O-bound, and the underlying RAID disk was averaging 100MB/s.

The advantages of each strategy can be summarised as follows:
\begin{itemize}
\item unordered: read Ids compress better as they are usually generated in an ordered way, which gets shuffled by the other strategies. This is a small gain.
\item LO sort: the dataset-end-pos file is not needed as the '\$' signs in the BWT end up in the same order as the original reads. This saves 7\% of the total size.
\item RLO sort: the BWT files get longer runs of identical letters, leading to a better compression rate (45\% of BWT size, 7\% of total size) and to faster search time (20\% faster).
\end{itemize}

The RLO sort achieves a 7\% reduction of the total size compared to the unordered strategy.
This comes at an extra cost in initialisation time, a stage where optimisations are still possible.
However, as our flow needs a single initialisation for an unlimited number of searches, the main benefits come at search time, where the I/O-bound search needs to go through a BWT which is 45\% smaller. Experiments confirm a 45\% faster search.

In the case of a server accessing files remotely, as described in the next section, another advantage lies in the fact that only BWT index files are kept cached in RAM. Having a BWT 45\% smaller means that a 45\% smaller RAM footprint is achieved with RLO.

It should be noted that the search-only time reported here corresponds to the first stage of reads extraction. The second stage (extension of BWT positions) is also accelerated by 45\% as it is based on the same BWT suffix extension algorithm, while the third stage (extraction of quality scores and read Ids from razip files) is unaffected.

\subsection{An \emph{in silico} pull-down experiment}
As a first application, we show how to use BEETL-fastq in a simple pull-down experiment. We use the gene RBM15B which is part of the
RNA-binding motif protein 15B. RBM15B is a relatively small gene consisting of a single exon with a total length of 6.6 Kb. We use it as a
illustrative example, but our method is not limited to single genes or exons.
We extract the list of non-overlapping 34-mers from the reference sequence. Using BEETL-fastq, we query the FASTQ files of NA12878 for all these k-mers
and retrieve all read pairs where at least one of the reads contains one of the k-mers from the list. We then re-align the read pairs using bwa-mem
to the Human reference genome and perform variant calling using Samtools~(\cite{Li2009}). 100\% of the retrieved read pairs align to the reference genome and 99.93\% 
of the reads align within the target region of RBM15B, which illustrates the high specificity of our approach.
Using whole-genome alignment and variant calling we identified a homozygous mutation ($T \rightarrow C$) in RBM15B. This mutation was confirmed to be part of the curated list of variants for NA12878 from the Genome in a Bottle project~\cite{Zook2014}.
Using the reads retrieved by BEETL-fastq and the standard workflow for variant calling using Samtools as described on the Samtools webpage, we correctly identify
this homozygous mutation. Retrieving the reads, aligning them and running samtools takes only minutes. The read alignment to the whole reference genome is by far the longest step and we
could reduce this further since we can estimate the approximate read reference location from the k-mer that was used to retrieve the read. Even unoptimized, the approach using BEETL-fastq 
is thus orders of magnitude faster than whole-genome alignment and of high interest when we are only interested in a small number of genes and their variants. 

\subsection{De-novo assembly of insertions}
One of the advantages of BEETL-fastq is its ability to extract not only the read containing the query k-mer but also its read partner from
the same DNA fragment. This opens new applications not accessible to other tools such as de-novo assembly of breakpoint regions but also
dissection of complicated breakpoint regions.
Unfortunately, the detection of large insertions is still difficult for the current generation of structural variant calling tools and 
comprehensive annotations even for well-characterized genomes such as NA12878 is difficult to obtain. 
To give an example that it is nevertheless possible to fully assemble large insertions using the query results of BEETL-fastq, we are
examining an insertion which we detected using an in-house structural variant calling pipeline as well by manual inspection. The insertion
occurs on chromosome 11 of NA12878 at position 5,896,446 and has a total length of 253bp. We extract the set of tiling 35-mers from the 
insertion as well as the k-mers crossing the insertion breakpoints.

\begin {figure}
{\scriptsize    %
$$          %
\includegraphics[bb = 0 0 520 253,width=85mm]{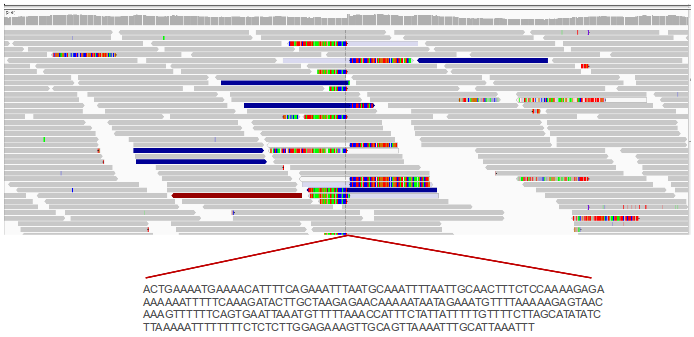}
$$
}
\caption{Fully assembly insertion sequence using reads extracted by BEETL-fastq}
\label{fig:InsertionNA12878}
\end{figure}

Using BEETL-fastq, we extract all read pairs in which at least one read contains an insertion k-mer. We then assemble these read pairs into
the velvet assembler~(\cite{Zerbino2008}) using standard parameters and a k-mer length of 31. As depicted in Fig~\ref{fig:InsertionNA12878},
the result is a single contig which we successfully aligned back to the reference genome and could confirm the insertion site.

\subsection{Structural variant breakpoints}

We use the genome NA12878 which is part of the above mentioned Utah pedigree. This genome was also sequenced as part of the 1000 Genomes
effort to characterise Human variation on a population scale~(\cite{1KG2012}) and an annotation consisting of 2702 deletions for NA12878, 
created using multiple SV callers, is publicly available~(\citep{Mills2010}). From this list, we removed the calls indicated as 
low quality and imprecise breakpoints in the annotation. This leaves us with a list of 1209 high-confidence deletions with precise 
breakpoints. The list also contains estimated genotypes for each variant.

We extracted the set of 34-mers crossing the deletion breakpoints and their reverse complements. We queried the fastq files of NA12878 for 
these k-mers to confirm the annotation by the 1000 Genomes consortium. We could retrieve reads for 1067 out of 1209 deletions (88.3\%). We
inspected some of the breakpoints for which we could not extract any k-mers and found these breakpoints have either ambiguous locations or
are closely adjacent to SNPs or small indels.

\begin {figure}
{\scriptsize    %
$$          %
\includegraphics[bb = 0 0 436 405, width=85mm]{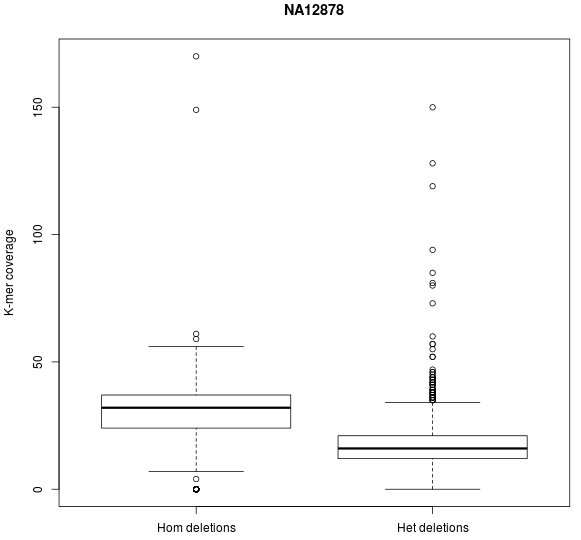}
$$
}
\caption{Read coverage of 1000 Genomes deletions in NA12878}
\label{fig:DeletionsKmersNA12878}
\end{figure}

As a next step, we compute the read coverage across the breakpoint from the reads extracted by BEETL-fastq and compared the results 
to the genotypes computed by the 1000 Genomes consortium. For heterozygous events, the coverage should be approximately half of the whole
genome coverage whereas for homozygous events breakpoint and whole genome coverage should be approximately the same. We found the breakpoint
coverages to be close to the expected values, a mean value of 17.2 for heterozygous and 29.52 for homozygous deletions. 
Fig~\ref{fig:DeletionsKmersNA12878} shows the spread of read coverages. There are some outliers which occur very frequently in the data. Closer
inspections reveals that they mostly map to short repeat instances.

After confirming that our genotype estimates match, we can now query these k-mers in other samples and thus test for the existence of these
deletions in other samples. This is a very powerful method and allows us to process large numbers of unaligned fastq files in a rapid manner. We can
also genotype and map variations across families and populations.

As an example, we examine the parents of NA12878, which we also sequenced as part of the Platinum Genome effort. They have the sample IDs 
NA12891 and NA12892. We query their sequencing reads stored as BWT-compressed fastq files and check if the estimated coverage match the 
inheritance pattern given the genotypes in NA12878. For instance, if NA12878 contains a homozygous deletion, we would expect to find
the corresponding k-mers in both parents. We found that the k-mer occurrences match this expected pattern in 99.96\% of all cases.

\section{Discussion}\label{sec:discussion}

Our focus here has been to represent the input data in a lossless fashion, but Table~\ref{fig:SortingStrategies} highlights that the relative storage needs of the sequences, quality scores and read IDs perhaps does not reflect their relative importance, since the sequences and their indexes take up only 17\% of the total size, barely more than the 14\% consumed by the read IDs.

There remains some scope for further lossless compression: in earlier work that focused exclusively on compression (\cite{CoxBauerJakobiRosone2012}), we achieved sub-0.5bpb on the sequences using 7-zip\footnote{\texttt{www.7-zip.org}, Igor Pavlov.}. Here, however, we sacrifice some compression for the faster decompression and thus faster search that our simple byte-code format gives us. 
The byte-coding can be further optimized and it may also be advantageous to switch between switch run-length encoding and naive 2-bits-per-base encoding on a per-block basis, choosing whichever of the two strategies best compresses each block.

However, to achieve significant further compression, some degree of lossy compression is likely to be necessary. Each of the three data streams are potentially amenable to this and our methods can of course still be applied to the resulting data. 

The free format of the read IDs limits our ability to comment generally on the prospects of compressing such data further. Nevertheless it could be argued that, for paired data, the most useful metadata to retain is which read is paired with which: this could be simply encapsulated in an array containing one pointer per read, consuming $O(\textrm{log} n)$ bits. 

The space taken up by the sequences themselves would be reduced by error correction, two possible strategies being the trimming of low-quality read ends (as demonstrated by \cite{CoxBauerJakobiRosone2012}) or a $k$-mer based approach such as Musket (\cite{Liu2013}).

Lastly, the majority of the archive's size is taken up by the quality scores. More recent Illumina data defaults to a reduced-representation quality scoring scheme that makes use of only 8 of the 40 or so possible quality values, but the PG data we tested still follows the full-resolution scoring scheme. The newer scheme would likely reduce the size of the scores by about half. We have also described a complementary approach that uses the $k$-mer context adjacent to a given base to decide whether its quality score can be discarded without likely detriment to variant calling accuracy.

\bibliographystyle{natbib}
\bibliography{BWT}

\begin{thebibliography}{}

\bibitem[Adjeroh {\em et~al.}(2008)Adjeroh, Bell, and
  Mukherjee]{bookBWTAdjeroh:2008}
Adjeroh, D., Bell, T., and Mukherjee, A. (2008).
\newblock {\em {The Burrows-Wheeler Transform: Data Compression, Suffix Arrays,
  and Pattern Matching}\/}.
\newblock Springer Publishing Company, Incorporated, 1st edition.

\bibitem[Bauer {\em et~al.}(2011)Bauer, Cox, and Rosone]{BauerCoxRosoneCPM11}
Bauer, M.~J., Cox, A.~J., and Rosone, G. (2011).
\newblock Lightweight {BWT} construction for very large string collections.
\newblock In {\em CPM 2011\/}, volume 6661 of {\em LNCS\/}, pages 219--231.
  Springer.

\bibitem[Bauer {\em et~al.}(2013)Bauer, Cox, and Rosone]{BauerCoxRosoneTCS2012}
Bauer, M.~J., Cox, A.~J., and Rosone, G. (2013).
\newblock Lightweight algorithms for constructing and inverting the {BWT} of
  string collections.
\newblock {\em Theoretical Computer Science\/}, {\bf 483}(0), 134 -- 148.

\bibitem[Burrows and Wheeler(1994)Burrows and Wheeler]{bwt94}
Burrows, M. and Wheeler, D.~J. (1994).
\newblock A block sorting data compression algorithm.
\newblock Technical report, DIGITAL System Research Center.

\bibitem[Cock {\em et~al.}(2010)Cock, Fields, Goto, Heuer, and Rice]{Cock2010}
Cock, P. J.~A., Fields, C.~J., Goto, N., Heuer, M.~L., and Rice, P.~M. (2010).
\newblock The sanger {FASTQ} file format for sequences with quality scores, and
  the {Solexa/Illumina} {FASTQ} variants.
\newblock {\em Nucleic Acids Research\/}, {\bf 38}(6), 1767--1771.

\bibitem[Cox {\em et~al.}(2012)Cox, Bauer, Jakobi, and
  Rosone]{CoxBauerJakobiRosone2012}
Cox, A., Bauer, M., Jakobi, T., and Rosone, G. (2012).
\newblock Large-scale compression of genomic sequence databases with the
  {B}urrows-{W}heeler transform.
\newblock {\em Bioinformatics\/}, {\bf 28}(11), 1415--1419.

\bibitem[Ferragina and Manzini(2000)Ferragina and Manzini]{Ferragina:2000}
Ferragina, P. and Manzini, G. (2000).
\newblock Opportunistic data structures with applications.
\newblock In {\em Proceedings of the 41st Annual Symposium on Foundations of
  Computer Science\/}, pages 390--398, Washington, DC, USA. IEEE Computer
  Society.

\bibitem[Langmead {\em et~al.}(2009)Langmead, Trapnell, Pop, and
  Salzberg]{bowtie2009}
Langmead, B., Trapnell, C., Pop, M., and Salzberg, S. (2009).
\newblock {Ultrafast and memory-efficient alignment of short DNA sequences to
  the human genome}.
\newblock {\em Genome Biology\/}, {\bf 10}(3), R25+.

\bibitem[Li and Durbin(2009)Li and Durbin]{bwa2009}
Li, H. and Durbin, R. (2009).
\newblock {Fast and accurate short read alignment with {B}urrows-{W}heeler
  transform}.
\newblock {\em Bioinformatics\/}, {\bf 25}(14), 1754--1760.

\bibitem[Li {\em et~al.}(2009)Li, Handsaker, Wysoker, Fennell, Ruan, Homer,
  Marth, Abecasis, Durbin, and Subgroup]{Li2009}
Li, H., Handsaker, B., Wysoker, A., Fennell, T., Ruan, J., Homer, N., Marth,
  G., Abecasis, G., Durbin, R., and Subgroup, . G. P. D.~P. (2009).
\newblock The sequence alignment/map format and samtools.
\newblock {\em Bioinformatics\/}, {\bf 25}(16), 2078--2079.

\bibitem[Liu {\em et~al.}(2014)Liu, Luo, and Lam]{Liu2014}
Liu, C.-M., Luo, R., and Lam, T.-W.~L. (2014).
\newblock {GPU}-accelerated {BWT} construction for large collection of short
  reads.
\newblock {\em {arXiv}: 1401.7457\/}.

\bibitem[Liu {\em et~al.}(2013)Liu, Schröder, and Schmidt]{Liu2013}
Liu, Y., Schröder, J., and Schmidt, B. (2013).
\newblock Musket: a multistage k-mer spectrum-based error corrector for
  illumina sequence data.
\newblock {\em Bioinformatics\/}, {\bf 29}(3), 308--315.

\bibitem[Mills {\em et~al.}(2011)Mills, Walter, Stewart, Handsaker, Chen,
  Alkan, Abyzov, Yoon, Ye, Cheetham, Chinwalla, Conrad, Fu, Grubert,
  Hajirasouliha, Hormozdiari, Iakoucheva, Iqbal, Kang, Kidd, Konkel, Korn,
  Khurana, Kural, Lam, Leng, Li, Li, Lin, Luo, Mu, Nemesh, Peckham, Rausch,
  Scally, Shi, Stromberg, Stutz, Urban, Walker, Wu, Zhang, Zhang, Batzer, Ding,
  Marth, {McVean}, Sebat, Snyder, Wang, Ye, Eichler, Gerstein, Hurles, Lee,
  {McCarroll}, and Korbel]{Mills2010}
Mills, R.~E., Walter, K., Stewart, C., Handsaker, R.~E., Chen, K., Alkan, C.,
  Abyzov, A., Yoon, S.~C., Ye, K., Cheetham, R.~K., Chinwalla, A., Conrad,
  D.~F., Fu, Y., Grubert, F., Hajirasouliha, I., Hormozdiari, F., Iakoucheva,
  L.~M., Iqbal, Z., Kang, S., Kidd, J.~M., Konkel, M.~K., Korn, J., Khurana,
  E., Kural, D., Lam, H. Y.~K., Leng, J., Li, R., Li, Y., Lin, C.-Y., Luo, R.,
  Mu, X.~J., Nemesh, J., Peckham, H.~E., Rausch, T., Scally, A., Shi, X.,
  Stromberg, M.~P., Stutz, A.~M., Urban, A.~E., Walker, J.~A., Wu, J., Zhang,
  Y., Zhang, Z.~D., Batzer, M.~A., Ding, L., Marth, G.~T., {McVean}, G., Sebat,
  J., Snyder, M., Wang, J., Ye, K., Eichler, E.~E., Gerstein, M.~B., Hurles,
  M.~E., Lee, C., {McCarroll}, S.~A., and Korbel, J.~O. (2011).
\newblock Mapping copy number variation by population-scale genome sequencing.
\newblock {\em Nature\/}, {\bf 470}(7332), 59--65.

\bibitem[Simpson and Durbin(2012)Simpson and Durbin]{Simpson2011}
Simpson, J.~T. and Durbin, R. (2012).
\newblock Efficient de novo assembly of large genomes using compressed data
  structures.
\newblock {\em Genome Research\/}, {\bf 22}(3), 549--556.

\bibitem[{The 1000 Genomes Project Consortium}(2012){The 1000 Genomes Project
  Consortium}]{1KG2012}
{The 1000 Genomes Project Consortium} (2012).
\newblock An integrated map of genetic variation from 1,092 human genomes.
\newblock {\em Nature\/}, {\bf 491}, 56--65.

\bibitem[Zerbino and Birney(2008)Zerbino and Birney]{Zerbino2008}
Zerbino, D.~R. and Birney, E. (2008).
\newblock Velvet: Algorithms for de novo short read assembly using de bruijn
  graphs.
\newblock {\em Genome Research\/}, {\bf 18}(5), 821--829.

\bibitem[Zook {\em et~al.}(2014)Zook, Chapman, Wang, Mittelman, Hofmann, Hide,
  and Salit]{Zook2014}
Zook, J.~M., Chapman, B., Wang, J., Mittelman, D., Hofmann, O., Hide, W., and
  Salit, M. (2014).
\newblock Integrating human sequence data sets provides a resource of benchmark
  snp and indel genotype calls.
\newblock {\em Nat Biotech\/}, {\bf 32}(3), 246--251.

\end{thebibliography}

\end{document}